\documentclass[11pt,twoside]{article}
\usepackage{atmp}
\copyrightnotice{2002}{6}{795}{826}	%% year, volume, first page, last page.
\begin{document}
\title{Boundary actions in Ponzano-Regge discretization, Quantum Groups
and $AdS_3$}
\url{gr-qc/0002092}		%% Everything after http://xxx.lanl.gov/
\author{Martin O'Loughlin}		%% Can have multiple \author, \address
\address{Spinoza Institute\\
        University of Utrecht\\
        Leuvenlaan 4\\
        Utrecht 3584 CE\\
        The Netherlands}
\addressemail{loughlin@sissa.it}	%% Under \address only, please!
% 
% \markboth{\it TITLE TO PROVE A POINT\ldots}{\it M.S. AUTHOR}

%\begin{document}
%\begin{titlepage}
%SPIN-2000/04, gr-qc/0002092

%\begin{center}
%{\LARGE{\sc Boundary actions in Ponzano-Regge discretization,}}\\
%{\LARGE{\sc Quantum groups and $AdS_3$}}\\
%\vskip .3in
%{\sc Martin O'Loughlin}\footnote{e-mail: loughlin@sissa.it}\\
%\vspace{.2in}
%{\it 
%\end{center}
\markboth{Boundary actions in Ponzano-Regge discretization ...}{Martin O'Loughlin}  
\begin{abstract}
Boundary actions for three-dimensional quantum gravity in the 
discretized formalism of Ponzano-Regge are studied with a 
view towards understanding the  boundary
degrees of freedom. These degrees of freedom postulated in the 
holography hypothesis are supposed to be characteristic of
quantum gravity theories. In particular it is expected that some 
of these degrees of freedom reside on black hole horizons.  
This paper is a study of these ideas in the context of a theory 
of quantum gravity that requires no additional structure such as
supersymmetry or special gravitational backgrounds. Lorentzian as well 
as Euclidean regimes are examined. Some surprising relationships
to Liouville theory and string theory in $AdS_3$ are found.
\end{abstract}

%\end{titlepage}
\cutpage

\section{Introduction}

This article presents the calculation, using continuum and lattice methods,
of boundary terms in 3-dimensional gravity. The gravity theory is presented
in first order Palatini form, this being a particular example of the general 
class of BF models \cite{BFpapers} 
as this is the most convenient presentation 
for deriving the discretization. We find a variety of boundary 
conditions, and discuss the significance of these for different types
of boundaries in space-time. 

The bulk theory of three-dimensional gravity is well known to be a 
topological field theory, however it is also well known that 
three-dimensional topological field theories can give rise to non-
topological boundary degrees of freedom, the classic example being the 
CS theory giving rise to a WZW model on the boundary \cite{emss}. 
In the case of three dimensional gravity with cosmological constant, 
one can utilize a trick that relates
the action to the difference of two CS actions, and then use
the standard CS-WZW relationship, however the actual boundary conditions
are a little more subtle. In three
dimensions this is relevant to the $AdS_3$ space, or more generally to
BTZ black hole solutions.

In this paper we wish to understand in the context of discretization of
quantum gravity the boundary degrees of freedom that correspond to 
black hole entropy.
This paper is directed towards a longer study of boundary terms in gravity 
theories, ultimately in $3+1$ dimensions, with the hope of understanding 
directly in a theory of quantum gravity, the possibile origin 
of holographic phenomena, and of the microscopic details of black hole 
entropy, in particular well out of the supersymmetric and extremal
limits which have been very well studied in the framework of 
string theory.

The actual type of discretization that we consider here is maybe at first
sight a bit unusual. The approach is originally due to Ponzano and Regge
\cite{PR} where they considered a simplicial decomposition of a three-manifold
and the path-integral is then defined as a summation over the 
possible sets of lengths of the edges of the dual lattice. The 
alternative of course is to fix the size of the simplices and to form the 
path integral by summation over possible simplicial decompositions. For the
major part of this paper, we will be discussing three dimensional models 
that have a topological invariance in the bulk 
and thus the fixed decomposition is 
somewhat innocuous but again the use of this simplicial decomposition
also for the boundary where in general we believe there are physical
degrees of freedom needs to be considered more cautiously. 
In addition we eventually need to extend our results to 
the realistic case of four-dimensional gravity where we do not even
have topological invariance in the bulk making things
more intricate though hopefully still manageable. 

We begin however, in the context of euclidean three-dimensional gravity
where already we find some interesting results concerning the boundary 
theories.
We will start off with a discussion of a discretization of the BF theory 
that corresponds to three-dimensional euclidean gravity in the framework 
of the Ponzano-Regge discretization, that is a discretization into 
tetrahedra with edges labelled by SO(3) spins, and each tetrahedron then
weighted in the path integral (sum) by the corresponding $6j$ symbol. 
From this discretization we can then derive a boundary action and will 
compare this to what we may expect from the corresponding BF theory. 
We in fact find that there are two simple types of boundary conditions,
one leads to a topological boundary theory and the other to a dynamical 
boundary theory. In addition we discuss mixed boundary conditions 
which are relevant for the boundary at infinity in $AdS_3$ for example. 
We discuss modifications to these boundary actions that
arise when one replaces the group $SO(3)$ with $SO(2,1)$ which would 
correspond to gravity with lorentzian signature. We also discuss
the regularization via quantum groups and find some interesting
relationships to work on string theory and $AdS_3$/CFT duality. 
 
Finally we make some suggestions for understanding black hole entropy
in this context and we discuss briefly the extension of these 
methods to four-dimensional quantum gravity. 

\section{Ponzano Regge from BF-theory}

We will now turn to a discretization of the BF representation of 
three-dimensional gravity and show how it leads to the Ponzano-Regge action.
The BF action is a generic action for a certain class of topological
field theories, \cite{BFpapers}. For three-dimensional gravity it actually 
corresponds to the Palatini first order action.
We will mostly use the $BF$ variables which are 
related to the gravity variables
via the dictionary;  $B = e$ is the dreibein 
and $F = R = d\omega + \omega\wedge\omega = dA + A^2$ is the curvature 
of the spin-connection $\omega = A$. 

The basic action for three-dimensional gravity in this first order 
formulation is then
\begin{equation}
S_{grav} = \int tr(e\wedge R),
\end{equation}
where $R$ is the curvature two form of a potential one-form $\omega$,
and $e$ is the dreibein. These fields transform under the action of
an $SO(3)$ gauge group. The invariance of the action consists
of a gauge transformation in $\omega$, $\omega\rightarrow h^{-1}\omega h + 
h^{-1}dh$, 
coupled with a local gauge rotation
$e \rightarrow h^{-1}eh$, and an additional invariance only acting 
on the dreibein (reparametrization)
under which $\delta e = d\chi + [\omega,\chi]$. The parameters of
these transformations are in $SO(3)$ and it's Lie algebra 
respectively. The first transformation expresses
the local lorentz invariance, and the second the diffeomorphism invariance. 
The theory as formulated is diffeomorphism invariant with no explicit
appearance of the metric in the action and thus topological. 
The constraint that the metric is torsion free, $de + \omega\wedge e = 0$,
in this first order form, arises from the $\omega$ equation of motion. 
The total group of local symmetry is $ISO(3)$ \cite{wittencs}.  

We will proceed now to a discrete formulation of three-dimensional gravity. 
We will carry out the discretization as a means of studying the continuum 
theory, however we would like to point out that in \cite{hooftpoly}
some arguments are given indicating that in three-dimensional gravity 
the space-time is necessarily discrete. Our study in fact also 
indicates another possibile method to prove that three-dimensional
gravity is discrete. 

One of the original motivations leading us to consider 
a discrete space-time approach to quantum gravity
is the following. We will throughout this paper
take the view that black holes in quantum gravity behave 
like quantum mechanical objects, and that this leads to unitarity in 
quantum gravity via some type of holographic mechanism \cite{Holography}. 
If one considers the black hole horizon to 
be a quantum object capable of storing and retransmitting information,
then one 
would imagine that this horizon follows a null or even time-like path in 
space-time and that the region inside the global horizon is not something 
that an outside observer can ever see or discuss. This is the 
view of black-hole complementarity developed to reconcile the 
apparent contradiction that unitary black hole evaporation implies
that observers outside the black hole view the physics of the 
horizon in a very different way to freely falling observers who fall 
into the horizon of a large black hole \cite{bhcomp}.
As such one may view the formation of a black - hole as 
the expansion of a planckian bubble in space-time to become macroscopic. 
Inside such a bubble there is nothing. Thus it seems 
necessary to think of 
the microscopic structure of space-time to be a collection 
of bubbles. As such there
is a discretization of space-time into units of size the Planck length. 

There are a variety of ways to approach the discretization of the BF theory 
in three-dimensions, although all constructions give the same final result. 
For other discussions of the approach that we present here 
see \cite{BFdiscrete,tftdisc}.
\begin{figure}
\centerline{\includegraphics{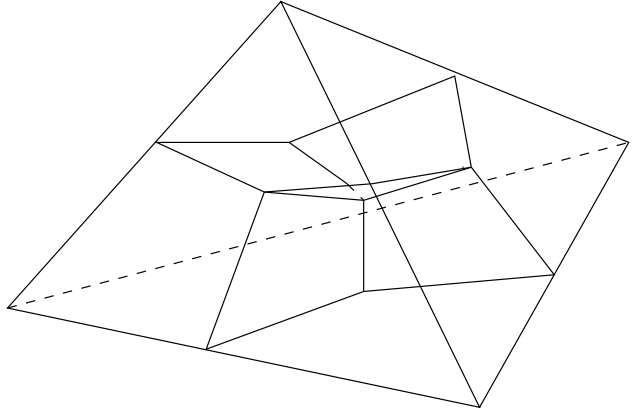}}
\caption{A tetrahedron and the the part of the dual lattice 
that it intersects.}
\end{figure}
%\EPSFIGURE[p]{tet.eps}{A tetrahedron and the the part of the dual lattice 
%that it intersects.}
The simplest approach to discretization is to formally carry out the path 
integral over the B-field, as it is simply a Lagrange mutiplier for the 
Einstein equation. The result is,
\begin{equation}
Z[M] = \int {\cal D}A \prod_x \delta(e^{F(x)})
\end{equation}
where $x$ are the co-ordinates on the closed manifold $M$, and the 
delta function is in the group manifold of $SO(3)$. The delta function can 
be rewritten using the identity 
\begin{equation}
\delta(gh^{-1}) = \sum_R \chi_R(g)\chi_R(h^{-1})
\end{equation}
where $g,h \in G$ and the sum is over all representations of the group G.
Using this identity we can write, 
\begin{equation}
Z[M] = \int {\cal D}A \prod_x \sum_j (2j + 1) \chi_j(e^{F(x)}).
\end{equation}
To make this expression tractable we now discretize the manifold $M$ by 
dividing it into tetrahedra. From this tetrahedral decomposition, we 
construct a dual discretization for which the vertices are at the centre
of the tetrahedra \cite{reisberg}, 
the edges pass between the centres of adjacent tetrahedra, 
and the faces are then bound by these edges and each dual face will be pierced
by precisely one edge of the original tetrahedral decomposition. In 
Figure 1 we show the part of the dual lattice that will live inside one of 
the original tetrahedra.

We now assign to every face of the dual lattice (that is every edge of the 
original lattice) a representation and to every edge of the dual lattice
a group element as shown in Figure 2. 
The product of the group elements around
a dual face is then the holonomy of that cycle and 
thus represents a discretization of the curvature. Denoting the discretization
of $M$ as $\Delta$ we can finally write,
\begin{equation}
Z[M,\Delta] = {\cal N}\int \prod_{e\in\Delta} \sum_{j_e} dU_e (2j_e + 1)
\chi_{j_e}(\prod_{\partial\tilde{e}}U)
\end{equation}
where ${\cal N}$ is a normalisation factor. 
In this expression, $e$ is an edge of the tetrahedral decomposition $\Delta$
and $\tilde{e}$ is the face dual to the edge $e$. 

\begin{figure}
\centerline{\includegraphics{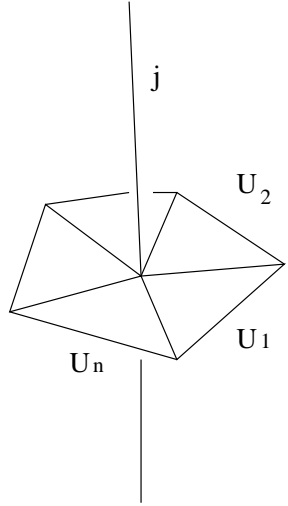}}
\caption{An edge of the lattice showing the corresponding dual 
face}
\end{figure}
%\EPSFIGURE[p]{dual.eps}{An edge of the lattice showing the corresponding dual 
%face}
To actually evaluate this expression we notice that the character can be 
written as a sum of products of the Wigner function $D^j_{mm'}(U)$ where 
$U$ is the group element corresponding to an edge of the dual graph,
\begin{equation}
\chi_j(\prod_{i=1}^nU_i) = \sum_{m_i}\prod D^j_{m_im_{i+1}}(U_i),
\end{equation}
where $m_{n+1} = m_n$. Then
for each edge of the dual graph there will appear in the integral over the 
corresponding group elements three Wigner functions which can be evaluated 
immediately using, 
\begin{equation}
\int dU D^{j_1}_{ll'}(U)D^{j_2}_{mm'}(U)D^{j_3}_{nn'}(U^{-1}) = 
\left (\begin{array}{ccc}
j_1 & j_2 & j_3\\    
l & m  & n
\end{array}\right )
\left (\begin{array}{ccc}
j_1 & j_2 & j_3\\    
l' & m' & n'
\end{array}\right )
\end{equation}
(For these and other angular momentum identities 
that we use below and for the definitions of the various symbols 
that we use, we recommend
that the reader refer to the very complete monograph \cite{vmk}).

Each tetrahedron thus will contribute two $3jm$ symbols for every face,
thus eight  $3jm$ symbols for every tetrahedron. Half of these are summed 
over angular momentum projections in pairs, one of the pair coming from
each of two tetrahedron with a common face and the orthogonality of the
$3jm$ symbols ensures that this term becomes the identity. The remaining 
expression is 
such that summing over the projection quantum number of the angular
momentum the four $3jm$ of a given tetrahedron gives 
a single $6j$ symbol using the identity,
\begin{eqnarray}
\lefteqn{\left\{\begin{array}{ccc}
j_1 & j_2 & j_3\\    
j_4 & j_5 & j_6
\end{array}\right\}=
\sum (-1)^{\sum j_i-\sum m_i} \left (\begin{array}{ccc}
j_1 & j_2 & j_3\\    
m_1 & m_2 & -m_3
\end{array}\right )\times}\notag\\
&\left (\begin{array}{ccc}
j_1 & j_5 & j_6\\    
-m_1 & m_5 & m_6
\end{array}\right )
\left (\begin{array}{ccc}
j_5 & j_3 & j_4\\    
-m_5 & m_3 & m_4
\end{array}\right )
\left (\begin{array}{ccc}
j_4 & j_2 & j_6\\    
-m_4 & -m_2 & -m_6
\end{array}\right )&
\end{eqnarray}
The final result for a closed manifold ${\cal M}$ and simplicial
decomposition $\Delta$ is (see \cite{BFdiscrete} or 
\cite{tftdisc} for more details), 
\begin{equation}
Z({\cal M},\Delta) = {\cal N}\sum_{j_e}\prod_{e\in\Delta}(2j_e + 1)
\prod_{t \epsilon \Delta}
(-)^{\sum_{i=1}^6 j_t^i} \left\{\begin{array}{ccc}
j_t^1 & j_t^2 & j_t^3\\    
j_t^4 & j_t^5 & j_t^6
\end{array}\right\}
\end{equation}

This answer is the path sum proposed by Ponzano and Regge to be a 
discretization of three dimensional quantum
gravity. In fact for a tetrahedron with 
edge lengths $l_i$, the corresponding weight is the 6j symbol with angular
momenta $j_i = l_i + \frac{1}{2}$. In the semi-classical limit \cite{PR} 
(large angular 
momenta and vanishing Planck length such that the combination $l_i\ell_{P}$ 
remains constant), a 6j symbol actually becomes the cosine of the
Regge action for the 
tetrahedron as a direct discretization of three dimensional gravity 
(the cosine arises as the BF theory path integral sums indiscriminately over
positive, negative and degenerate values for $B$). The Regge action is 
the direct discretization of the Einstein-Hilbert action \cite{Regge}.
\begin{eqnarray}
S_{Regge} &=& \sum_{h,k =1}^4 (j_{hk} + \frac{1}{2}) \theta_{hk}\\
\left\{\begin{array}{ccc}
j_1 & j_2 & j_3\\    
j_4 & j_5 & j_6
\end{array}\right\} &\simeq& \frac{1}{\sqrt{12\pi V}} \mbox{cos}
\left( S_{Regge} + \frac{\pi}{4}\right)
\end{eqnarray}
$\theta_{hk}$ is the angle between the normals to adjoining faces and 
$j_{hk}$ is the
length of the edge common to the two faces labeled by $h$ and $k$.
The extra factor of half between the edge length and corresponding angular
momentum is for consistency in this semi-classical limit and we can intuitively
justify it by noting that the length
of the angular momentum vector for the representation of spin $j$ is actually
$\sqrt{j(j+1)}$ which becomes $j + \frac{1}{2}$ in the limit of large angular 
momentum. 

\subsection{Symmetry and normalization}

The above expression for the discretized path sum is not quite complete. We
have ignored the fact that there could in principle be some normalization
factor in front of the sum, and in fact one would hope that there is such 
a factor simply because the sum itself is divergent. One can simply choose 
a normalization factor to subtract the divergence, however it is interesting
to see how the divergence arises. This was already analyzed in the original
paper of Ponzano and Regge, and the reader should look there for the details. 
In short, one takes the Biedenharn-Elliot (BE) identity (Appendix A.), 
which relates a product 
of three $6j$ symbols summed over one angular momentum, to a product of two
$6j$ symbols without summation. Geometrically this corresponds to taking 
three tetrahedra joined together along a common edge, and each with a face 
in common with two of the others. Removing the common edge (the sum in the 
BE identity) leaves one with two tetrahedra sharing one common face. 
Using the orthogonality of the $6j$ symbols, one can change this 
identity to one that relates a single tetrahedron to four tetrahedra 
formed by introducing an additional  
vertex at the centre of the original tetrahedron.
The identity is in Appendix A for the interested reader. The important 
point is that there is an infinite factor
\begin{equation}
\Lambda (R) = \lim_{R\rightarrow\infty}\sum_{j=0}^R (2j+1)^2,
\end{equation}
for every vertex of the simplicial decomposition. 

Therefore we see that an infinite factor of this form must be added to 
the denominator of the path sum to regularize it, and that there is one 
such factor for every vertex in the triangulation. The actual
normalization factor is then 
\begin{displaymath}
{\cal N} = \Lambda(R)^{-N_v}
\end{displaymath}
where $N_v$ is the number of vertices in the discretization. 

In addition this 
discussion has shown us that the path sum is actually invariant under
the two transformations derived from the BE identity. These two 
transformations are known as Pachner moves \cite{pachner} and these are the 
discretized version of diffeomorphisms. We have therefore learnt that the
path sum thus defined (in particular with the regularization discussed)
is diffeomorphism invariant in discretized form just as the $BF$ theory 
was before the discretization.

\subsection{Regularization}

The full path sum is then, 
\begin{eqnarray}
\lefteqn{Z({\cal M},\Delta) = }\\
&&\lim_{R\rightarrow\infty}\Lambda (R)^{-N_0}
\sum_{j_e}^R\prod_{e\in\Delta}(2j_e+1)
\prod_{t\in\Delta}
(-)^{\sum_{i=1}^6 j_t^i} \left\{\begin{array}{ccc}
j_t^1 & j_t^2 & j_t^3\\    
j_t^4 & j_t^5 & j_t^6
\end{array}\right\}\notag
\end{eqnarray}
where $N_0$ is the number of vertices in $\Delta$. In this form however it 
is still not very practical for calculating. There exists a different 
regularization that involves a q-deformation of $so(3)$ due to 
Turaev and Viro \cite{TV}. 

The path sum is
\begin{eqnarray}
\lefteqn{Z_{TV}({\cal M},\Delta) = }\\
&&\Lambda_q^{-N_0}
\sum_{j_e=0}^{\frac{k-1}{2}}\prod_{e\in\Delta}[2j_e+1]_q
\prod_{t\in\Delta}
(-)^{\sum_{i=1}^6 j_t^i} \left[\begin{array}{ccc}
j_t^1 & j_t^2 & j_t^3\\    
j_t^4 & j_t^5 & j_t^6
\end{array}\right]_q\notag
\end{eqnarray}
where 
\begin{align}
\Lambda_q &= -\frac{2k}{(q - q^{-1})^2}\\
[n]_q &= \frac{q^n - q^{-n}}{q - q^{-1}}
\end{align}

The parameter of the quantum deformation is a root of 
unity $q = e^{\pi i/k}$ and the 
sum is regularized as the representations of $U_q(so(3))$ 
involve angular momenta
only in the range $0\dots (k-1)/2$ so the path 
sum now involves all finite sums and $\Lambda(R)$ has been replaced
by $\Lambda_q$ which is clearly finite. 

The semi-classical
limit of the $q-6j$ symbol indicates that the q-deformed path sum is related 
to quantum gravity in three dimensions with a positive cosmological constant. 
The limit must be carried out in a way that as the angular momentum become
large, correspondingly also $k$ must go to infinity. The limit is \cite{mizta},
\begin{equation}
\left\{\begin{array}{ccc}
j_1 & j_2 & j_3\\    
j_4 & j_5 & j_6
\end{array}\right\}_q \simeq \frac{1}{\sqrt{12\pi V}} \mbox{cos}
\left( S_{Regge} - \frac{4\pi^2}{k^2}V + \frac{\pi}{4}\right)
\end{equation}
and we see in particular that the limit which makes contact with the 
semi-classical physics is the limit in which the cosmological constant
goes to zero. Note that this cannot be derived directly from the action
\begin{equation}
S_{\Lambda} = \int tr(BF + \Lambda B^3)
\end{equation}
by any simple generalization of the discretization carried out above, as 
the non-linearity in $B$ does not allow us to easily integrate over $B$ to 
get a simple expression involving the curvature $F$. It would be very 
interesting to find a derivation of the TV path sum from the 
discretization of the path integral for $S_\Lambda$.

For simple manifolds this sum can actually be evaluated giving 
the Turaev-Viro invariants that are important for the understanding of 
the topology of three-manifolds. The restriction on angular 
momentum in the quantum group representations 
is the same as that which must be imposed on string states in $AdS_3$. 
We will take another look at the q-deformed action and limits thereof
after we have discussed the boundary discretization and will find that
in the context of gravity in $AdS_3$ there may indeed 
be a deeper meaning to this regularization. 

It is also interesting to consider 
the relationship between this construction of three-dimensional
gravity using a quantum deformation and studies of  
quantum doubles of groups \cite{Bais}. In this article one has a different
type of quantum group that does not have a fixed deformation 
parameter. It is used for the discussion of multi-particle states in 
three-dimensional gravity. Each particle, creates a localized 
source of curvature, and in general the space is conical at infinity. 
It is amusing to notice that for the Chern-Simons description of 
three-dimensional gravity, at zero cosmological constant one uses
the group $ISO(3)$ \cite{wittencs}, but at non-zero cosmological 
constant, one finds instead the group $SU(2)\times SU(2)$ with 
the level of the Chern-Simons theory related to the curvature.
Going to the multi-particle Fock space in three-dimensions
means that we are allowing variable localized curvature depending 
upon the location and mass of the particle sources. We 
find that the group $ISO(3)$ is replaced by ${\cal D}(SU(2))$ \cite{Bais}
but now with 
no additional parameter, indicating perhaps that all values of 
curvature are possibile depending on the number and mass of particles 
present. This relationship deserves to be studied in more detail as it 
indicates a possible second quantization that involves also the 
cosmological constant. 

\section{Boundaries}

Let us consider the general variation of the $BF$ action for a manifold 
with boundary, (other work on this subject can be found in the papers
\cite{ooguri,oogsas,BJT}). 
\begin{equation}
\delta S_{BF} = \int_M tr(\delta B F + \delta A (dB + AB + BA)) -
\int_{\partial M} tr(B \delta A)
\end{equation}
We see then that the field equations are not effected by the presence of 
the boundary provided that the variation of $A$ is zero on the boundary. 
The path integral in the presence of the boundary will then be a function 
of the boundary value of the spin connection. 

We have another choice, which corresponds to the BF theory with a boundary 
term
\begin{equation}
S = S_{BF} + \int_{\partial M} tr(BA)
\end{equation}
The variation of $S$ is now
\begin{equation}
\delta S = \int_M \mbox{``equations of motion''}
+ \int_{\partial M} tr (\delta B A)
\end{equation}
and therefore the boundary condition must be that the variation of $B$ is
zero on the boundary, and the path integral will now be a function of the 
boundary metric. 

The first boundary condition of fixed spin connection on the boundary 
actually gives rise to a topological field theory on the manifold plus
boundary.
The second boundary condition is Dirichlet on the metric, and this does
not give rise to a topologically invariant boundary action.

In a study of asymptotic symmetries in 
three-dimensional gravity \cite{BHliou},
it was shown that with appropriate boundary conditions one can also find 
a Liouville theory on the boundary at infinity of $AdS_3$ space. 
Such boundary conditions formulated in terms of the metric and connection 
are actually mixed boundary conditions, and we will give more details of
how these work below. 

The  continuum boundary action can be easily derived by 
following a construction similar to that used in
\cite{emss} where the WZW-CS relationship was discussed  in some detail. 
First we consider the boundary condition $\delta\omega = 0$ for which there is
no additional boundary term. To examine the boundary theory we will insert
the solutions to the bulk equations and then examine the action of 
the gauge symmetries of the theory in the presence of the boundary. For
the WZW-CS relationship the boundary degrees of freedom arise precisely
because the bulk gauge symmetry is only a global symmetry on the boundary
thus the breaking of the gauge symmetry by the presence of the boundary 
gives rise to new degrees of freedom. 

The bulk equations of motion are solved by 
\begin{equation}
A = -dUU^{-1},
\end{equation}
\begin{equation}
B = UdVU^{-1}.
\end{equation}
Substituting these solutions into the action we find of course that it 
vanishes identically as $R=0$. The gauge variation of the action is 
identically zero for the local lorentz invariance, but the diffeomorphism
transformation has in principle an additional boundary term equal to
\begin{equation}
\delta_{diff} I = \int_{\partial M} tr(\chi R)
\end{equation}
which we see also vanishes as $R=0$ from the equations of motion
(modulo some topological issues regarding the extension of a flat
connection to a boundary of given topology). 
This boundary action is that of an obviously topological two-dimensional 
field theory in agreement with the proof by 
Ooguri and Sasakura \cite{oogsas} that with the $\delta\omega = 0$ boundary condition 
the path sum is a topological invariant not just of the bulk but of the
bulk plus boundary theory. 

For the $\delta e = 0$ boundary condition we must add to this result
the boundary term $\int tr(e\omega)$. The boundary condition now seems to 
indicate that the boundary metric is important in the path sum, in fact
the path sum will now be a function of the boundary triangulation. Again
the solutions to the bulk equations will be inserted into the action,
the bulk again giving zero contribution but the boundary now gives a non zero 
contribution equal to
\begin{equation}
S = -\int_{\partial M} tr(dVU^{-1}dU).
\end{equation}
Furthermore the gauge transformations now give rise to non-trivial 
boundary terms, 
\begin{equation}
\delta_{gauge} S = \int_{\partial M} tr(\Lambda dB)
\end{equation}
\begin{equation}
\delta_{diff} S = -2\int_{\partial_M} tr(\chi A^2)
\end{equation}
We can see from these variations that the symmetry of the boundary theory
is significantly smaller than that of the bulk theory. In fact we must have
$\Lambda$ constant for the gauge transformation to vanish and also 
$\chi = 0$.
Therefore the boundary theory has no diffeomporphism invariance, and 
is invariant only under global lorentz transformations. 

Finally we can consider the boundary conditions used in \cite{BHliou}
which are related to three dimensional gravity with cosmological constant.
To do this we make a small deviation into the Chern-Simons representation
of three-dimensional gravity with cosmological constant. 
Our action is then,
\begin{equation}
S_{BF} = \int_{\cal M} tr(B F + \Lambda B^3)
\end{equation}
We make the change of variables, 
\begin{equation}
A^{\pm} = \frac{1}{2}(B\sqrt{-3\Lambda} \pm A)
\end{equation}
and we then find that $S_{BF}$ becomes the difference of two 
Chern-Simons theories plus an additional boundary term.  
\begin{equation}
S_{BF} = \frac{1}{\sqrt{-3\Lambda}}
\int_{\cal M} (CS[A^+] - CS[A^-]) + 
\frac{1}{\sqrt{-3\Lambda}}\int_{\partial {\cal M}}tr(A^+A^-)
\end{equation}
We see here that the level of the Chern-Simons theory is inversely 
proportional to the square root of the cosmological constant, and also
that if we started with a $BF$ action with no boundary term, then
after the change of variables we have a boundary term that is of a mixed
form, rather than of the form $tr(AB)$. This is due to the fact 
that using the variables $A^\pm$ we can consider boundary conditions that 
would be mixed boundary conditions when expressed in terms of 
the variables $A$ and $B$. Indeed, if we add
$\frac{1}{2}\int tr(AB)$ 
to the $BF$ action then following the construction of
\cite{emss} one finds that in the Chern-Simons variables the action 
factorizes into two pieces that represent a pair of chiral WZW theories. 

The boundary conditions now imply restrictions on a combination of the 
metric and connection. It is precisely this setup that was shown
to arise for the boundary at infinity of $AdS_3$ in the work
of Brown and Henneaux and afterwards Coussaert, Henneaux and van Driel
\cite{BHliou}. 
The boundary theory is actually a Liouville theory. Note that to discuss 
this case in the discretized framework we really need to use the quantum 
group representations as it is only then that ones sees a cosmological
constant in the semi-classical limit. The discretized boundary theory will 
turn out to be very similar to a discretization of Liouville theory. We will
show how this relationship arises in more detail once we have set up the 
formalism for the quantum discrete boundaries. One may already worry here that
we are trying to construct some triangulation of Liouville theory in the
strongly coupled phase and it is well known that for Euclidean surfaces
such theories have very non-continuum like phases. A discussion of these
problems and arguments for better behaviour in the lorentzian case 
are in \cite{lordiscr}.

\subsection{Quantum discrete boundaries}

From the bulk calculation of the discretized path sum, we saw that 
for every face of the simplicial decomposition, there are two $3jm$ symbols.
Indeed if we consider a single tetrahedron as a discretization of
the three-dimensional ball then it has a weight,
%\begin{eqnarray} 
%\lefteqn{(-1)^{\sum_{i=1}^6 j_i}
%\left\{\begin{array}{ccc}
%j_1 & j_2 & j_3\\    
%j_4 & j_5 & j_6
%\end{array}\right\}
%(-1)^{(i_2 + i_3)}
%\left(\begin{array}{ccc}
%j_1 & j_2 & j_3\\    
%i_1 & -i_2 & -i_3
%\end{array}\right)\times}\\
%&&(-1)^{(k_1 + k_3)}
%\left(\begin{array}{ccc}
%j_1 & j_5 & j_6\\    
%-k_3 & -k_1 & k_2
%\end{array}\right)
%(-1)^{(l_1 + l_3)}
%\left(\begin{array}{ccc}
%j_4 & j_2 & j_6\\    
%-l_1 & l_2 & -l_3
%\end{array}\right)
%\left(\begin{array}{ccc}
%j_4 & j_5 & j_3\\    
%m_2 & m_3 & m_1
%\end{array}\right)\notag
%\end{eqnarray}
\begin{eqnarray}
&(-1)^{\sum_{i=1}^6 j_i +(i_2 + i_3 + k_1 + k_3 + l_1 +l_3)}
\left\{\begin{array}{ccc}
j_1 & j_2 & j_3\\
j_4 & j_5 & j_6
\end{array}\right\}
\left(\begin{array}{ccc}
j_1 & j_2 & j_3\\
i_1 & -i_2 & -i_3
\end{array}\right)\times&\\
&
\left(\begin{array}{ccc}
j_1 & j_5 & j_6\\
-k_3 & -k_1 & k_2
\end{array}\right)
\left(\begin{array}{ccc}
j_4 & j_2 & j_6\\
-l_1 & l_2 & -l_3
\end{array}\right)
\left(\begin{array}{ccc}
j_4 & j_5 & j_3\\
m_2 & m_3 & m_1
\end{array}\right)&\notag
\end{eqnarray}

Joining now an additional tetrahedron to one of the faces of this
tetrahedron, we get another decomposition of the three-ball. On the internal 
face there is now a $3jm$ symbol coming from each of the tetrahedra, but we
must now sum over the angular momentum projections assigned to the internal 
faces (now identified of course).
Using the orthogonality identity for a pair of $3jm$ symbols
\begin{equation}
\sum_{m_i}
\left(\begin{array}{ccc}
j_1 & j_2 & j_3\\    
m_1 & m_2 & m_3
\end{array}\right)
\left(\begin{array}{ccc}
j_1 & j_2 & j_3\\    
m_1 & m_2 & m_3
\end{array}\right) = 1
\end{equation}
we see that the internal $3jm$ symbols vanish and we are left in the path sum
with a $6j$ symbol for every bulk tetrahedron, and a $3jm$ symbol
for every boundary face. 
\begin{figure}
\centerline{\includegraphics{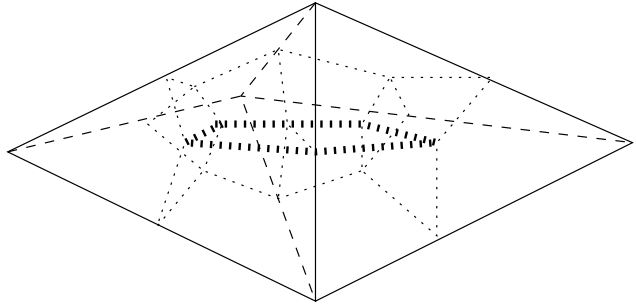}}
\caption{A pair of boundary triangles and their tetrahedra. 
Dotted lines are the dual lattice. Bold dotted lines
highlight a dual face cut by the boundary.}
\end{figure}
%\EPSFIGURE[p]{cut.eps}{A pair of boundary triangles and their tetrahedra. 
%Dotted lines are the dual lattice. Bold dotted lines
%highlight a dual face cut by the boundary.}
In general the integral that gave rise to this pair of $3jm$ symbols
was along the link of the dual lattice that passes from the centre of one
tetrahedra to the centre of an adjacent one 
piercing one and only one face. In the 
presence of a boundary only half of this integral is carried out, from the
centre of the tetrahedron to the face and this integral gives rise to a $3jm$
symbol for the bulk $6j$ symbol and additional single $3jm$ symbol
for the boundary face. The new feature that has given rise to the boundary
weights for the boundary faces is that now we do not have an entire dual face,
but rather the dual face is cut in half by the presence of the boundary
as shown in figure 3. 
 
We therefore need to also consider the group elements 
that live on the edge of the 
dual face that is exposed by the boundary. For the bulk path-sum described in 
the previous section the connection was integrated away. 
Now due to the exposed
dual faces, we have a boundary dependence on the connection that we may or may
not integrate over depending upon the boundary conditions chosen. 
In figure 4 we have labelled
one such edge from $X$ to $Y$ with its weight $D^j_{mn}(U)$. For the 
boundary conditions that correspond
to the action with no boundary term, that is the $\delta A = 0$ conditions, 
we are instructed to keep the connection
fixed on the boundary, and thus we must not integrate over the boundary 
values of $U$. We thus find a network with trivalent vertices, each vertex
is weighted by a $3jm$ symbol, and the vertices are tied together by the 
matrix elements of the corresponding group 
elements. The one and two tetrahedra
path sums above easily generalize by gluing faces of tetrahedra together and 
using the orthogonality condition giving one the general expression for 
a simplicial decomposition with boundary. 

\begin{eqnarray}
\lefteqn{Z({\cal M},\partial {\cal M},\Delta, \partial\Delta) =}\notag\\
&&{\cal N}\sum_{\{j_e\}}\prod_{e\in\Delta}(2j_e + 1)
\prod_{t \in \Delta}
(-)^{\sum_{i=1}^6 j_t^i} \left\{\begin{array}{ccc}
j_t^1 & j_t^2 & j_t^3\\    
j_t^4 & j_t^5 & j_t^6
\end{array}\right\}\times\\
&&\sum_{\{m^i_f\}}\prod_{f\in\partial\Delta}(-)^{\frac{1}{2}\sum m^i_f}
\left(\begin{array}{ccc}
j_f^1 & j_f^2 & j_f^3\\    
m_f^1 & m_f^2 & -m_f^3
\end{array}\right)
\prod_{e\in\partial\tilde{\Delta}}D_{m_e,m_e'}^{j_e}(U_e)\notag
\end{eqnarray}

\begin{figure}
\centerline{\includegraphics{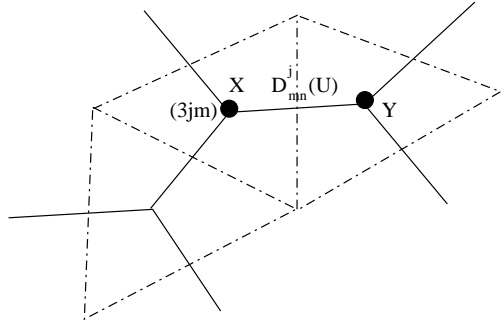}}
\caption{Boundary discretization arising from the boundary of 
bulk tetrahedra and the dual lattice (trivalent graph). }
\end{figure}
%\EPSFIGURE[p]{bdry.eps}{Boundary discretization arising from the boundary of 
%bulk tetrahedra and the dual lattice (trivalent graph). }

In this expression the normalization factor is the usual one mentioned above
and $\tilde{\Delta}$ is the dual lattice. 
The summation is over the angular momenta assigned to 
edges in the bulk and the boundary, and over the angular momentum 
projections assigned to each triangular face of the boundary.
For the situation where the group representations summed over are those 
of the quantum group this is precisely the bulk plus boundary 
action derived by Ooguri and Sasakura 
\cite{oogsas}, where they show that the Hilbert space of the $TV$ theory is 
equivalent to that of a pair of Chern-Simons theories
for which the boundary state is described by Wilson
lines joined by trivalent vertices with an identical structure to that 
derived above. We would also like to note that this path sum (for all 
boundary group elements $U$ equal to the identity element) is the same as that 
derived in \cite{ccm}. In contrast to our present approach, in that paper
the boundary action was derived purely on the grounds of topological 
invariance. 

The boundary term $\int_{\partial M} tr(BA)$ 
required for the $\delta B = 0$ boundary 
conditions when discretized becomes, 
\begin{equation}
exp(\int_{\partial M} tr(\bar{B}A)) = \chi_j(U) = \sum_mD^j_{mm}(U)
\end{equation}
where $\bar{B}$ refers to the boundary value of $B$. In this expression 
the dreibein $\bar{B}$ is replaced by its discretized representation that 
being the length of the corresponding edge of the boundary of the original
lattice, and the connection is represented by $U$ which is the gauge 
field assigned to the link of the boundary of the dual lattice that is 
dual to the edge where $\bar{B}$ resides. The partition function 
is in this case a function only of $\bar{B}$. We must multiply the path 
sum derived above for the $\delta A = 0$ boundary conditions by this 
additional term, remove the sum over the boundary values of the spins $j_f$,
and integrate over $U$ to derive the final path sum for Dirichlet boundary
conditions in the metric. The integral of importance is that over $U$ 
and is
\begin{equation}
\int dU D^j_{mm'}(U) D^k_{nn}(U) = \frac{1}{2k+1} \delta_{jk}\delta_{mn}
\delta_{m'n}
\end{equation}
Inserting this into the path sum gives the final result for fixed metric 
boundary conditions, 
\begin{eqnarray*}
\lefteqn{Z({\cal M},\partial {\cal M},\Delta, \partial\Delta) =}\\
&&{\cal N}\sum_{\{j_e\in\Delta '\}}\prod_{e\in\Delta '}(2j_e + 1)
\prod_{t \in \Delta}
(-)^{\sum_{i=1}^6 j_t^i} \left\{\begin{array}{ccc}
j_t^1 & j_t^2 & j_t^3\\    
j_t^4 & j_t^5 & j_t^6
\end{array}\right\}\times\\
&&\sum_{\{m^i_e\}}\prod_{f\in\partial\Delta}(-)^{\frac{1}{2}\sum m^i_e}
\left(\begin{array}{ccc}
j_e^1 & j_e^2 & j_e^3\\    
m_e^1 & m_e^2 & -m_e^3
\end{array}\right)
\end{eqnarray*}
where $\Delta'$ signifies the lattice without boundary components.
The sum is now only over the angular momentum in the interior edges. 
The integral over $U$ on the boundary has now fixed the angular 
momentum projections to be associated to edges of the boundary 
triangulations, rather than with faces as for $\delta A= 0$.
Thus we see that the action is quite similar to that for the $\delta A = 0$ 
boundary conditions except that now the boundary values of the angular
momenta are fixed corresponding to the fixed boundary metric. Note that in
this path sum the factors of $(2j + 1)$ are absent for edges that lie in 
the boundary due to the restriction in the product over edges to 
$\Delta^\prime = \Delta - \partial\Delta$. These factors are important
in the angular momentum identities that one uses to prove topological 
invariance, thus indicating that for this choice of boundary conditions
there is no topological invariance on the boundary agreeing with our 
continuum analysis. 

In the path integral for a fixed boundary metric, one would expect that in
the quantum gravity there would be a need to sum over all possible 
boundary configurations that give a discretization of the continuum
boundary metric. A construction of such a type will be seen to 
be necessary for a calculation of black hole 
entropy in this discretized setup. 
In general before fixing boundary conditions we have the expression 
for $\delta A = 0$ without summation over angular momenta and without
the integral over the boundary gauge connection. We need to understand
what boundary conditions will allow calculations relevant to 
black hole physics, and also what representations of the (quantum) 
group one must include in this summation. 
The representations and boundary conditions
will be discussed in the final section
when we consider the construction for lorentzian metrics. 
Furthermore, we need to know how to implement the boundary conditions
that give Liouville theory in our path sum construction. 

\subsection{Two-dimensional discrete path sums}

We want to show a point of contact between 
our calculations and discrete TFT's in two dimensions. 
For $\delta A = 0$ everything is topological and there is an easy way to get
a two dimensional TFT from this theory. 
In $R^3$ take a thickened wall and 
remove the bulk tetrahedra using the various Pachner moves in the bulk and 
on the boundary. The final result will be just two dimensional, but in some 
sense a double layer as the two faces will both carry their own $3jm$ symbols. 
The two dimensional action that one finds by this procedure is,
\begin{eqnarray}
\lefteqn{Z(\Sigma, U) = \sum_{\{j_e\}}\prod_{e\in\Sigma}(2j_e+1)
\sum_{\{m_f^i, m_f^i\prime\}}\prod_{f\in\Sigma}
(-)^{\frac{1}{2}\sum (m_f^i + m_f^i\prime)}\times}\notag\\
&\left(\begin{array}{ccc}
j_f^1 & j_f^2 & j_f^3\\    
m_f^1 & m_f^2 & -m_f^3
\end{array}\right)&
\left(\begin{array}{ccc}
j_f^1 & j_f^2 & j_f^3\\    
m_f^1\prime & m_f^2\prime & -m_f^3\prime
\end{array}\right)\times\\
&&\prod_{e\in\tilde{\Sigma}}D_{m_e,n_e}^{j_e}(U_e)
D_{m_e\prime,n_e\prime}^{j_e}(U_e)\notag
\end{eqnarray}
This is indeed a two-dimensional TFT, invariant under 
two-dimensional pachner moves and similar actions have been
studied in a collection of works \cite{twdltft,ccm}.

For $\delta B = 0$ we cannot actually remove all the bulk 
tetrahedra, as the removal process that one uses for the 
totally topological situation of $\delta A = 0$ relies 
heavily on the topological invariance of the boundary theory and 
in particular on the elementary shelling operations. 
We can however take a limit that is inspired by the bulk boundary 
correspondence of the AdS/CFT conjecture \cite{adscft}. To do this we imagine
that we take a semi-classical limit of the bulk action leaving the boundary 
angular momentum fixed. The relevant limit of the bulk $6j$ symbols that 
have a face edge or vertex on the 
boundary were already studied in the original
article of Ponzano and Regge. The interesting thing that we find is that 
the boundary answer depends crucially on the asymptotic properties 
of the manifold. This sort of behaviour is maybe not a surprise
as it is precisely such a dependence in the AdS/CFT correspondence that 
accounts for the simplicity of the near horizon limit in the AdS case.
For asymptotically flat spaces however the action is not 
expected to be similar to the CFT as it will live
on a null surface rather than on a time-like surface and the 
asymptotic group of symmetries will be smaller. 

The limits of $6j$ symbols in which only some of the angular momentum are
taken to be large are of two basic types. The first involves removing one
vertex to infinity, and thus the three edges connected to that 
vertex become large, while the three vertices that form the remaining face
stay fixed, this face then is a triangle of the boundary configuration. 
The result is thus the $3jm$ symbol of the remaining face, 
where the pairwise differences between the large angular momenta make the 
$m$ quantum numbers in this $3jm$ symbol and we thus find an answer
similar to that which we derived from the BF theory, a pair of $3jm$
symbols on the boundary. The answer is,
\begin{eqnarray}
\lim_{R\rightarrow\infty}\lefteqn{\left\{\begin{array}{ccc}
j^1 & j^2 & j^3\\    
j^4 + R & j^5 + R & j^6 + R
\end{array}\right\}}& \notag \\
&\simeq (-1)^{\sum_{i=1}^3 j^i + 2\sum_{i=4}^6 j^i}
(2R)^{-\frac{1}{2}}
\left (\begin{array}{ccc}
j^1 & j^2 & j^3\\    
j^5 - j^6 & j^6 - j^4 & j^4 - j^5
\end{array}\right )&
\end{eqnarray}

The other possibility corresponds
to holding the length of one edge fixed, this representing a tetrahedra 
that has only an edge in contact with the boundary. In this case one still can 
do one of two things with the remaining angular momenta. One can take the 
angular momentum on the unique edge that does not touch our chosen edge
to also be fixed, and the other four go to infinity. Or one can take 
all five to be large. This is where the dependence on the large scale
asymptotics of the space have an effect. If for instance in the 
euclidean case we are considering a boundary that is a sphere in $R^3$, 
then clearly we must take the limit where all five other angular momenta 
become large. On the other hand, if the boundary is a plane in $R^3$ 
then one need take only four angular momenta to infinity, the other two
corresponding to opposite edges of the tetrahedra remain fixed. The expressions
for these limits contain additional dependence on parameters of the limiting 
process and can be found in appendix B. 

The expressions for the path sums in these limits 
are relatively complicated. It is interesting to note that 
the answer for this ``near-boundary'' limit, is basically the 
two-dimensional double $3jm$ symbol action derived above for purely 
topological boundary conditions, however, with some additional 
structure depending upon the asymptotic behaviour of the 
space-time. In the next section for null surfaces in 
Lorentzian manifolds we will find that the semi-classical
limit leads to an hypothesis that simplifies the boundary discretization
considerably. 

\section{Lorentzian manifolds, Liouville theory and string theory on 
$AdS_3$}

If we replace the $SO(3)$ of the euclidean construction with $SO(2,1)$ then 
the representation theory becomes somewhat more complicated, and all 
limits of the corresponding angular momentum coupling coefficients 
in the various representations have not been 
fully studied. However, the original large 
angular momentum limit of Ponzano and Regge has also been carried out 
for the discrete series of representations 
in the non-compact case \cite{semlortz}. The result 
is basically the same as for the compact group apart from the 
fact that the angles are now hyperbolic given by the boost required to 
take the normal to a face into the normal of an adjoining face.
In the limit of large angular momentum we have 
\begin{equation}
\left\{\begin{array}{ccc}
j_1 & j_2 & j_3\\    
j_4 & j_5 & j_6
\end{array}\right\} \simeq \frac{1}{\sqrt{12\pi |V|}} \mbox{cos}\,\phi
\,exp\left( -\left|\sum_{h<k}j_{hk}\Theta_{hk}\right|\right)
\end{equation}
In this expression, $\Theta_{hk}$ is the angle between the faces $h$ and $k$,
\begin{displaymath}
\Theta_{hk} = \mbox{cosh}^{-1}(n_h.n_k)
\end{displaymath}
and $n$ is the unit normal to the corresponding face. 
Thus as one face becomes null, the corresponding normal will also become
null, and the angle that this face makes with the three neighbouring 
faces becomes infinite. The exponential in the weight for the 
tetrahedron implies that the corresponding angular momentum must
be zero or that two of the sides of the triangle are of equal 
length and the corresponding angles are of opposite sign. 
Therefore the only configurations that can contribute 
have  equilateral triangles and isosceles triangles where the 
short edge has length $1/2$ corresponding to zero angular momentum. 
The equilateral triangles must have all zero angular momentum labels
and thus have all sides of length $1/2$. 
So in a path sum involving all discretizations with a 
given boundary metric the path sum is dominated by discretizations with 
boundary triangles that have all lengths equal to $1/2$. This is modified
then by collective structures built from isosceles triangles. It is 
clear that the configurations involving isosceles triangles must be 
collective, as the presence of an isosceles triangle, implies also that 
neighbouring triangles are isosceles, and so on, until the structure
closes again. An example of such a collective structure is shown 
in figure 5. These structures are reminiscent of macroscopic
loop operators in the matrix models of dynamical triangulations
\cite{macloops}. 

We should note here that we have been a bit incautious regarding the 
order of limits. We took large $j$ and then interpreted the expression
for small $j$. There is indirect evidence that the result is sensible  
and we will discuss our reasoning below. The precise calculation that 
one needs to do is to take the null boundary limit of the 
quantum $6j$ symbol in a similar way to the original limits studied
by Ponzano and Regge. 

\begin{figure}
\centerline{\includegraphics{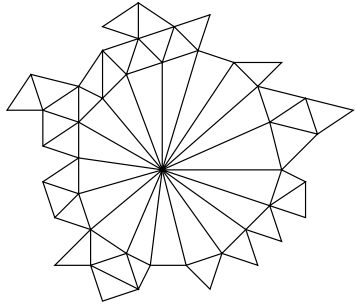}}
\caption{An example of part of a horizon configuration
showing an isosceles ``excitation''.}
\end{figure}
%\EPSFIGURE[p]{coll.eps}{An example of part of a horizon configuration
%showing an isosceles ``excitation''.}

Thus we get a picture of horizon states in 
discretized quantum gravity and this is a 
positive step towards a micrpscopic understanding of black hole entropy.
In 't Hooft's discussions of horizon states \cite{transpl} one finds
similarly a special role for the low angular momenta, $l=0,\pm\frac{1}{2}$
when the horizon at fixed Rindler time  is represented as a 
collection of discretized line segments labelled 
by angular momenta of $SO(2,1)$.
Also in the Ashtekhar approach to quantum gravity, the entropy calculations
indicate that entropy is derived from contributions only from the lowest spin
states on the horizon \cite{ashentropy} and similarly in 
the paper \cite{btzent}.

It is interesting to reflect upon the meaning of the boundary action. 
If we assume that the semi-classical limit of the Clebsch-Gordon 
coefficients for $U_q(sl(2))$ for the discrete representations
are an analytic continuation of those for $U_q(so(3))$ then we should
find a negative cosmological constant. 
For the situation of $2+1$ gravity in a space of constant negative 
curvature, one finds as mentioned above that the boundary theory is
a Liouville theory. Furthermore from recent work on Liouville theory
\cite{DO,ZZ,pontesch}
it is known that the representations of the Virasoro algebra 
that arise in the N-point functions, involve the quantum group
$U_q(sl(2,R))$.
In the string theory picture of $AdS_3$/CFT duality \cite{ads3string}
the mass cut-off on angular momentum representations is also the 
same as that which arises in the discrete representations 
of $U_q(sl(2,R))$. Beginning as we did from the PRTV
(Ponzano-Regge-Turaev-Viro) construction, it appears that we have arrived 
at almost the same conclusion. 
Note though that in the PRTV construction, after changing 
to a Lorentzian space-time signature it is not necessary that the 
representations are identical to those used for the 
Euclidean geometries. Maybe one should sum(integrate) over the continuous
representations that arise in the Liouville approach 
for the boundary at infinity. On the other hand, for the null boundary at 
a global horizon, it is not so clear how to proceed, however some interesting 
insight will come from a comparison of our boundary action 
and string theory on $AdS_3$ \cite{wip}. 
If we had the Clebsch-Gordon coefficients for the
discrete series of $U_q(sl(2))$ we could also explicitly calculate
the weight of a ``macroscopic loop'' configuration 
and make a direct comparison with
the macroscopic loop wavefunctions calculated for example in \cite{macloops}.
The Clebsch-Gordon coefficients are known for the continuous series
\cite{pontesch} and for these one should be able to directly compute the null 
boundary limit. 

The representation theory of non-compact quantum groups is still very much 
under development, see \cite{stein1, pontesch}. 
Also a discussion on the relationship between
strings in $AdS_3$ and quantum groups can be found in \cite{JR}.
There are  representations of $U_q(sl(2))$ that are discrete, and agree 
basically with the discrete ones for the compact group, and give a cut-off
in the path sum, see also \cite{stein1} for a few more details on these.
The other representations are those that arise 
from the quantum group representation 
of the Virasoro algebra of the Liouville cft at $c > 1$. These are 
similar to the continuous representations of $sl(2,R)$. 

We can already make some speculative remarks derived 
from studies of string theory and continuum gravity in $AdS_3$
\cite{ads3string}. One 
can study the various physical excitations in this space-time both from 
the perspective of the space-time and that of the string theory.
In the space-time picture, one finds a $c>1$ Liouville theory, and 
indeed if one considers a non-critical string theory with target 
equal to $AdS_3$ then again one will find the world-sheet theory 
also to be Liouville with $c>1$. The 
states that arise are classified by quantum group representations 
\cite{DO,ZZ,pontesch}. However the representations that
arise are not those that we are using in the  Turaev-Viro path sum. 
This strongly suggests that an extension of the PRTV
(Ponzano-Regge-Turaev-Viro) construction to include the representations
of $U_q(sl(2))$ that arise in the Liouville theory corresponds to 
extending the quantum gravity path sum, to a string field theory
path sum (albeit with a fixed topology for the target manifold).
Liouville theory at $c=1$ also appears in the context of $AdS_5$ 
compactifications although in this case the theory appears  
as a consequence of $SU(2)$ group factors in the internal space \cite{RDJat}. 
It would be interesting to find connections between this structure and
the Liouville theory that is naturally present for $AdS_3$ string 
compactifications. 

The Liouville theory on the boundary cylinder at infinity 
for gravity in $AdS_3$ has a central charge 
\begin{equation}
c = 1 + 6(b + 1/b)^2
\end{equation} 
where $b\in R$ or $|b| = 1$ and the correlation 
functions of this theory are constructed from the Clebsch-Gordon
coefficients of the quantum group $U_q(sl(2))$ where 
\begin{equation}
q = e^{(i\pi b^2)}.
\end{equation}
The cosmological constant of the $AdS_3$ space is proportional
to $b^4$. In turn, the cosmological constant that arises in PRTV is 
proportional to $1/k^2$, where the deformation for the Turaev-Viro 
quantum group is given by,
\begin{equation}
q = e^{(\frac{i\pi}{k})}.
\end{equation}
Clearly  $b^2 \sim 1/k$ and thus the groups that arise in the 
two approaches are indeed identical deformations 
of $sl(2)$. Thus the group and
deformation parameter agree in a manner which supports the conjecture that
the quantum group that arises in the Regge calculus is the same as 
that of the Liouville theory on the boundary. However,
the representations that arise in the Liouville theory have 
angular momentum in $\frac{Q}{2} + i{\cal R}$, while those in PRTV
are identical to those that arose for $U_q(so(3))$ with angular 
momentum running from 0 to $\frac{k-1}{2}$. Of course
once we changed from Euclidean to Lorentzian discretizations, the question 
already arose as to which representations one should sum over and now
we see that the answer to this question may have deeper significance. 

Actually one can make the relationship between our discrete boundary
action involving the quantum group and the perturbation theory 
of the Liouville theory on the cylinder more concrete in a very 
geometrical manner by examining the perturbative expansion of 
the Liouville theory on a cylinder (corresponding to the boundary
of $AdS_3$). 
Write the path integral with sources and charges for all Liouville
vertex operators in selected representations. 
Use bootstrap to argue that all vertices can be reduced to cubic and 
recall that the cubic vertex for the Liouville theory is given
precisely by the Clebsch-Gordon coefficient of the quantum group $U_q(sl(2))$
\cite{pontesch}. Furthermore the propogator of the 
perturbative expansion of the Liouville theory is the Wigner coefficient
$D^j_{mn}$ of the corresponding representations. Such Feynman
diagrams correspond precisely to the dual lattice with weights as
derived in the previous section and as shown in Figure 3.
Geometrically all genus zero amplitudes correspond to one of 
our quantum group boundary terms in structure but with a sum over
representations different from those used by Turaev-Viro. 

If one considers the dual lattice to the boundary triangulation, one finds
a trivalent graph that lives on the boundary of the manifold, being 
one of the Feynman diagrams discussed above. At any given
time-slicing this will look like a collection of particles with mass 
given by their spin, and as this gas evolves there are interactions 
coming from the trivalent graph. Thus one can make a proposal 
for calculating the entropy using a system of particles 
making a gas. The $XXZ$ spin chain is a possible starting point
for such a calculation. This model is a chain of spins the solution
to which involves the quantum group $U_q(sl(2))$ and which 
is related to Liouville theory for $c=1$ and $c > 25$ and also possibly
for all $c > 1$ \cite{fadtirk}. Within this framework we should be 
able to formulate the explicit calculation that is necessary to calculate
the entropy of the boundary theory and thus black hole entropy. 
\begin{figure}
\centerline{\includegraphics{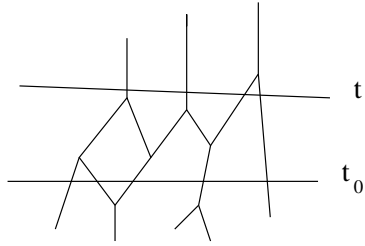}}
\caption{Two time slices of the boundary gas}
\end{figure}
%\EPSFIGURE[p]{gas.eps}{Two time slices of the boundary gas}
The Feynman diagrams of
the boundary action describe the time-evolution of the gas (Figure 6). 
For a null 
boundary the gas will be non-relativistic whereas for a time-like boundary 
the gas will be relativistic. 
In the case of a null boundary these representations will become more
restricted and the  three point interaction implies
that during the evolution of the gas one has both creation and annihilation
of particles. This may even imply some sort of dissipation in the null case.
Other works arguing
for dissipative behaviour for a theory describing a black hole horizon have 
appeared in \cite{hooftdiss, mempw}. From a deeper understanding of
this gas one should be able to directly calculate the entropy
and thus the black hole horizon entropy.

Another consideration that we have not addressed directly
but that has already arisen
a few times in our discussions, and also 
one that is intimately related to the calculation of the entropy is
the following. Without a boundary, it was clear that the prescription
of Ponzano and Regge to hold fixed the simplicial decomposition
was already sufficient due to the topological nature of the theory. 
Now in the presence of a boundary it is possible that 
one really needs to sum over the boundary triangulations. The bulk 
theory is topological and is insensitive to how one describes the sum in 
detail, however we expect some dynamics on the boundary. This indicates
the possibility of extending the path-sum to dynamical triangulations. This
sounds like trouble as such triangulations give rise to the matrix model 
of Liouville and for $c>1$ these models are badly behaved with very rough 
surfaces dominating the path sum. However, discretizations 
for lorentzian manifolds 
have been studied in 
\cite{lordiscr} where the authors have shown that when the 
simplicial decompositions are restricted by the requirement 
of a causal structure, 
the phases of the dynamical triangulations are well behaved 
involving smoother
surfaces than in the Euclidean setup.  The Haussdorf 
dimension in particular remaining $d_H = 2$ rather than becoming fractal and 
equal to 4 as it does in the Euclidean case. Indeed, our 
work also implies that 
there is another possibly interesting type of dynamical 
triangulation, where the "causal" structure is that implied by the 
constraint that the surface be not Lorentzian, but null. 
It would be interesting
to study in the context of \cite{lordiscr} null dynamical triangulations, 
the results of which investigation would certainly shed light on 
the dynamics of black hole horizons in quantum gravity. 

\subsection{$3+1$ dimensions}

For $3+1$ dimensions we now have some intuition for how to approach the
discretization. The simplices will be labelled by $SO(3,1)$ representations.
We can write the boundary path sum including boundaries following more or less
the same philosophy as above. In this case from the beginning it seems that 
we probably need to consider dynamical triangulations as otherwise we will 
end up with a topological bulk theory rather than a theory containing
also gravitational dynamics. Various versions of discretizations
of four-dimensional Lorentzian manifolds have been studied as for example
in \cite{fourdtop,BClrntz}. 
Furthermore the semi-classical limit of the $15j$ symbols that arise in 
these bulk path-sums, has been studied in \cite{barrwill} 
with results agreeing with
the Regge discretization once more. 
We expect for null boundaries
also in $3+1$ dimensions that some restrictions will be placed on the 
representations arising and that one will probably again find some sort of 
three-dimensional dynamical triangulation describing the behaviour of the 
horizon. 
From the work in \cite{fordiscr} it has been shown that also
for three-dimensional lorentzian dynamical triangulations, the branched polymer
and crumpled phases, can not be reached 
leaving hope that such a system will have
a nicely behaved continuum phase transition. 
It would be interesting to also look at three dimensional
null dynamical triangulations to see if the causality restrictions
on triangulations introduce some regulator of the geometries. 
The way to proceed is we believe clear. One must determine the representations
that are important for the theory that is being investigated, and then
one must look at various limits of the $j$-symbols.
\\[0.5cm]

\appendix

{\bf\Large Appendices}
\section{Angular momentum identities}

The Biedenharn-Elliot identity relates the $6j$ symbols associated to 
two different ways of combining nine angular momenta. The sum on the right
hand side is replaced by a product on the left. Geometrically this 
identity is represented by the diagram shown. 

\begin{eqnarray}
\lefteqn{\left\{\begin{array}{ccc}
j_7 & j_8 & j_9\\    
j_5 & j_1 & j_4
\end{array}\right\}
\left\{\begin{array}{ccc}
j_7 & j_8 & j_9\\    
j_6 & j_2 & j_3
\end{array}\right\}=}\\
&&\sum_X (-1)^{(\sum_i j_i + X)}
\left\{\begin{array}{ccc}
j_1 & j_2 & X\\    
j_3 & j_4 & j_7
\end{array}\right\}
\left\{\begin{array}{ccc}
j_3 & j_4 & X\\    
j_5 & j_6 & j_8
\end{array}\right\}
\left\{\begin{array}{ccc}
j_5 & j_6 & X\\    
j_2 & j_1 & j_9
\end{array}\right\}\notag
\end{eqnarray}

Using the orthogonality for a pair of $6j$ symbols this identity can be 
rearranged as discussed in the text, up to an infinite multiplicative factor.
The regularized version of this identity as first given in Ponzano and Regge
\cite{PR}, is

\begin{eqnarray}
\lefteqn{\left\{\begin{array}{ccc}
j_1 & j_2 & j_3\\
j_4 & j_5 & j_6\end{array}\right\}
= \lim_{R\rightarrow\infty}\Lambda(R)^{-1}\sum_{j_7\cdots j_{10}}
\prod_{i = 7\cdots10}(2j_i + 1)\times}\\
&&\left\{\begin{array}{ccc}
j_1 & j_2 & j_3\\
j_7 & j_8 & j_9\end{array}\right\}
\left\{\begin{array}{ccc}
j_6 & j_5 & j_1\\
j_8 & j_9 & j_{10}\end{array}\right\}
\left\{\begin{array}{ccc}
j_4 & j_2 & j_6\\
j_9 & j_{10} & j_7\end{array}\right\}
\left\{\begin{array}{ccc}
j_3 & j_5 & j_4\\
j_{10} & j_7 & j_8\end{array}\right\}\notag
\end{eqnarray}

Another useful identity for understanding the relationship between bulk 
and boundary transformations is, 
\begin{eqnarray}
\lefteqn{\sum_{m_3}(-1)^{j_3 - m_3}
\left(\begin{array}{ccc}
j_1 & j_2 & j_3\\
m_1 & m_2 & m_3\end{array}\right)
\left(\begin{array}{ccc}
j_3 & j_4 & j_5\\
-m_3 & m_4 & m_5\end{array}\right)
=}\\&&\sum_{j,m}(-1)^{j-m}
\left(\begin{array}{ccc}
j_1 & j_5 & j\\
m_1 & m_5 & m\end{array}\right)
\left(\begin{array}{ccc}
j & j_4 & j_2\\
-m & m_4 & m_2\end{array}\right)
\left\{\begin{array}{ccc}
j_2 & j_4 & j\\
j_5 & j_1 & j_3\end{array}\right\}\notag
\end{eqnarray}
The geometrical meaning of the left side is simply a pair of 
adjoining boundary faces. The right hand side involves the gluing
of two faces of an additional tetrahedron to the original pair
of faces resulting in a new pair of boundary triangles. This 
results in a $2\leftrightarrow 2$ Pachner transformation in 
two-dimensions. Using the orthogonality of the $3jm$ symbols one can
rewrite this equation to give the algebraic representation of 
the $3\leftrightarrow 1$ transformation. 

\section{Limits of $6j$ symbols}

Here are the $2+2$ and $2+1+1$ limits of Ponzano-Regge. 
We will use the following labelling for the tetrahedron (Figure 7). For the 
$2+2$ limit, we shift $b,c,e,f$ by $R$ and take $R$ to be much larger
than all of $a\dots f$. In the figure this limit corresponds 
to keeping the segments $[1,2]$ and $[3,4]$ of fixed length while
all other edge lengths go to infinity. 
We can consider $a = \mbox{length}[1,2]$ to 
be the edge of the tetrahedron that lies in the boundary. 

The answer is then, 
\begin{eqnarray}
\lefteqn{\left\{\begin{array}{ccc}
a & b+R & c+R\\
d & e+R & f+R\end{array}\right\}\simeq(-1)^{a+d+min(b+e,c+f)}}\notag\\
&&\left[\frac{(a-b+c)!(a-e+f)!(d-e+c)!(d-b+f)!}
{(a+b-c)!(a+e-f)!(d+e-c)!(d+b-f)!}\right]^{\frac{1}{2}\mbox{sign}
(c+f-b-e)}\times\notag\\
&&\frac{(2R)^{-|b+e-c-f|-1}}{|b+e-c-f|!}\left[1 + O(R^{-2})\right]
\end{eqnarray}

\begin{figure}
\centerline{\includegraphics{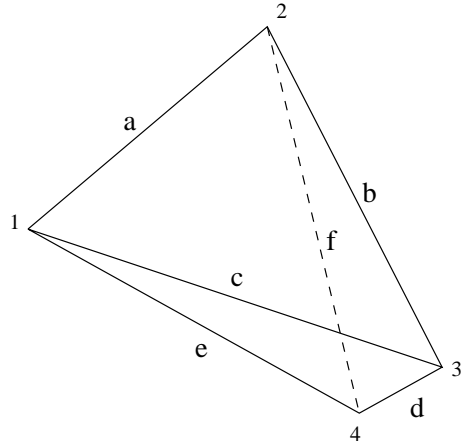}}
\caption{Some limits of tetrahedra}
\end{figure}
%\EPSFIGURE[p]{twotwo.eps}{Some limits of tetrahedra}
For the $2+1+1$ limit, we take $e=b+\delta$ and $f=c+\delta^\prime$,
where now $d,\delta,\delta^\prime$ are all large, though small with 
respect to $a,b,c$. This corresponds to keeping only 
the segment $d$ of fixed length and all other edges to infinity.
The one edge of small size is then $d$ and in the 
text this is the edge that lies in the boundary of the manifold, all
other edges in this case being internal. The final answer is,
\begin{equation}
\left\{\begin{array}{ccc}
a & b & c\\
d & b+\delta & c+\delta^\prime \end{array}\right\}\simeq
\frac{(-1)^{a+b+c+\delta+\delta^\prime}}{\left[12\pi V\right]}
\mbox{cos}\,(t-\frac{1}{4}\pi),
\end{equation}
where
\begin{equation}
t = \Omega - (a + b + c + \delta + \delta^\prime - \frac{1}{4})\pi
\end{equation}
and $\Omega$ is the Regge action for the tetrahedron. 

In this limit, the dependence
on the asymptotic structure of the space enters as the angle that remains
in the final expression is the angle between the edges $[2,3]$ and $[1,3]$ 
or equivalently between $[1,4]$ and $[2,4]$. These angles enter the 
expression for the limit through the Regge action.
If for instance the boundary is on 
a sphere of finite volume, then as one takes this limit a tetrahedron with 
one edge stuck on the sphere boundary, these angles will go to infinity. 
If the boundary is planar in flat space then the angles will go to zero 
and we go back to the $2+2$ result.
\\[0.5cm]

\centerline{\bf Acknowledgments}

The author would like to thank G.Arcioni, M. Blau, G. Thompson, 
G. 't Hooft and K. Krasnov
for useful conversations during the various stages of this work. This work 
was commenced during the Extended Workshop in String Theory at the Abdus Salam
ICTP in Trieste in the summer of 1999.  The author is supported by the 
Pionier Programme of the Netherlands Organisation for Scientific 
Research (NWO).

\end{document}